\documentclass{article}
\usepackage[preprint]{spconf}
\usepackage{amsmath,graphicx,color}
\usepackage{subcaption}
\usepackage{multicol}
\usepackage{amsmath,bm}
\usepackage{booktabs}
\usepackage{paralist}
\usepackage{hyperref}
\usepackage{cite}
\usepackage{enumitem}
\usepackage[makeroom]{cancel}

\copyrightnotice{\copyright\ IEEE 2018}
\DeclareMathOperator*{\argmax}{arg\,max}

\def\vec#1{\ensuremath{\boldsymbol{{#1}}}}
\def\mat#1{\vec{#1}}

\newcommand\textss[1]{\textsuperscript#1}

\title{A comparison of techniques for language model integration in encoder-decoder speech recognition}
\name{Shubham Toshniwal\textss{1}, Anjuli Kannan\textss{2}, Chung-Cheng Chiu\textss{2}, Yonghui Wu\textss{2}, Tara N Sainath\textss{2}, Karen Livescu\textss{1} }
\address{
\textss{1}Toyota Technological Institute at Chicago, USA\\
\textss{2}Google Inc., USA\\
  {\small \tt {\{shtoshni, klivescu\}@ttic.edu}, \{anjuli, chungchengc, yonghui, tsainath\}@google.com}
  }

\begin{document}
\ninept
\maketitle
\begin{abstract}
Attention-based recurrent neural encoder-decoder models present an elegant solution to the automatic speech recognition problem. 
This approach folds the acoustic model, pronunciation model, and language model into a single network and requires only a parallel corpus of speech and text for training.
However, unlike in conventional approaches that combine separate acoustic and language models, it is not clear how to use additional (unpaired) text.
While there has been previous work on methods addressing this problem, a thorough comparison among methods is still lacking.
In this paper, we compare a suite of past methods and some of our own proposed methods for using unpaired text data to improve encoder-decoder models. 
For evaluation, we use the medium-sized Switchboard data set and the large-scale Google voice search and dictation data sets.
Our results confirm the benefits of using unpaired text across a range of methods and data sets.
Surprisingly, for first-pass decoding, the rather simple approach of shallow fusion performs best across data sets.
However, for Google data sets we find that cold fusion has a lower oracle error rate and outperforms other approaches after second-pass rescoring on the Google voice search data set.
\end{abstract}


\section{Introduction}
Attention-based recurrent neural encoder-decoder models provide an elegant end-to-end framework for speech recognition, machine translation, and other sequence transduction tasks~\cite{Chan16, Wu16}.  
In automatic speech recognition (ASR), the model folds the traditionally separately learned acoustic model, pronunciation model, and language model (LM) into a single network that can be trained end-to-end.  The encoder maps the input speech to a sequence of higher-level learned features, while the decoder maps these higher-level features to output labels with assistance from the attention mechanism that provides an alignment between speech and text. 
The model can be learned end-to-end and requires just paired speech and text data.
Encoder-decoder models for speech recognition have become quite popular recently and perform competitively on a number of ASR tasks~\cite{Zeyer18, Chorowski16, Chiu18}.

While end-to-end training offers several advantages, it also restricts the training data to have both input and output sequences, 
for example the paired speech and text data in the case of speech recognition. 
Conventional ASR models leverage a separate LM trained on all available text, which can be orders of magnitude larger than just the transcripts of transcribed audio. Decoder of an encoder-decoder model is exposed only to the audio transcripts.

Previous work addressing the issue of utilizing unpaired text has proposed ways of integrating an external pretrained LM, trained on all of the text data, with the ASR model~\cite{Gulcehre15, Sriram17, Kannan18}.
The main LM integration approaches from past work have been referred to as shallow~\cite{Gulcehre15}, deep~\cite{Gulcehre15}, and cold fusion~\cite{Sriram17}. 
The three approaches differ in two important criteria:
\begin{itemize}
    \item \textit{Early/late model integration:} At what point in the ASR model's computation should the LM be integrated? 
    In deep and cold fusion, the external LM is fused directly into the ASR model by combining their hidden states, 
    resulting in a single model with tight integration.  In contrast, in shallow fusion the LM and ASR model remain separate and only their scores are combined, similarly 
    to an ensemble.  The shallow fusion score combination is also similar to the interpolation of acoustic and language models done in traditional ASR. 
    \item \textit{Early/late training integration:} At what point in the ASR model's training should the LM be integrated?  
    Deep and shallow fusion use a late integration where both the ASR and LM models are trained separately and then combined, while cold fusion uses the external pretrained LM model from the very start of the ASR model training.  An important point is that early training integration approaches are computationally costlier if either of the two models is frequently changing.
\end{itemize}

A thorough comparison between these LM integration techniques is, to the best of our knowledge, currently lacking.
In this paper, we compare the three fusion approaches mentioned above on (a) the medium-sized Switchboard data set~\cite{Godfrey92} and (b) the large-scale Google voice search and dictation data sets used in~\cite{Chiu18}. 
Our aim is to shed light on how the LM integration approaches compare, as well as how they scale up with data size.
We also propose some novel LM integration approaches and compare them against the three prior fusion approaches on Switchboard.

Our results show that almost all of the LM integration approaches improve over a baseline encoder-decoder model for all data sets, confirming the benefit of utilizing unpaired text.  
We also make several other findings: (a) the rather simple approach of shallow fusion works best for first-pass decoding on all of our data sets,
(b) our best proposed approach performs similarly to deep and cold fusion on the Switchboard data set,  
(c) deep fusion doesn't scale well, obtaining no or negligible gains over baseline for large-scale Google data sets, and 
(d) cold fusion produces high-quality and diverse beam outputs resulting in lowest oracle word error rate on Google data sets and edges ahead when coupled with second-pass LM rescoring on Google voice search.
\section{Related Work}
Previous work on using unpaired text for encoder-decoder models can be categorized along two major themes: \\

\noindent \textbf{Using an external language model}\\
This approach consists of training an external LM on the unpaired text and integrating it into the encoder-decoder model, which is the focus of this paper.
    An early study along these lines was by Gulcehre {\it et al.}~\cite{Gulcehre15}, who proposed the shallow and deep fusion methods in the context of neural machine translation (NMT) models.  In that work both shallow and deep fusion improved performance, with deep fusion somewhat outperforming shallow fusion, especially for low-resource language pairs.
    Another previous work in context of NMT models by Ramachandran {\em et al.}~\cite{ramachandran2017unsup} proposed initializing the lower layer of both encoder and decoder with separate pretrained LMs followed by joint training 
    using both language modeling and machine translation losses.
    Shallow fusion has largely been the method of choice for ASR~\cite{Chorowski16, Battenberg17, Kannan18, Zeyer18}, getting significant performance gains, although in some cases with slight modifications to the decoding objective function
    \cite{Chorowski16, Battenberg17, Kannan18}. 
    Cold fusion, a modification of deep fusion, was proposed for ASR by Sriram {\it et al.}~\cite{Sriram17}. 
    This work found that, on medium-scale data sets of $\sim$300-400K training utterances, cold fusion outperforms deep fusion, especially in a cross-domain setting, but did not compare with shallow fusion.  
    None of these studies compared all three fusion approaches.\\
    
\noindent \textbf{Generating paired data from unpaired text}\\
A second line of research is to use unpaired text to synthetically generate matching input sequences, thus expanding the paired data set.  In machine translation this process of generating paired data from monolingual data is referred to as backtranslation---that is, generating source-language text from unpaired target-language text---and has been used in the context of neural machine translation by Sennrich {\it et al.}~\cite{Sennrich16NMT}. 
    The directly analogous approach for ASR would be to use text-to-speech synthesis to generate speech from unpaired text. 
    The complexity of the text-to-speech (TTS) task means that there has been limited work exploring the use of 
    speech generated from unpaired text, often in limited settings~\cite{Tjandra17}. 
    A workaround of ``translating" the text to phoneme sequences and using the resulting paired data in a multitask learning setup has been explored by Renduchintala {\it et al.}~\cite{Renduchintala18}.

\section{Model}
Our model is based on the Listen, Attend and Spell (LAS) attention-based encoder-decoder ASR model proposed by~\cite{Chan16}.  We begin by reviewing this model, and then describe the techniques we consider for LM integration with the LAS model. 

\subsection{LAS Model}
The LAS model consists of three components: an {\em encoder}, a {\em decoder}, and an {\em attention network} which are trained jointly to
predict the output sequence. The transcription can be decoded as a sequence of graphemes/characters, wordpieces, or words from a sequence of acoustic feature frames. 
Based on the recent success of wordpiece-based models in a variety of ASR tasks and machine translation tasks~\cite{Sennrich16, Wu16, Chiu18, Zeyer18}, we choose wordpieces as output unit in all our models.

The encoder consists of a stacked (bidirectional) recurrent neural network (RNN)~\cite{Schuster97} which reads in acoustic features $\vec{x} = (\vec{x}_1, \cdots , \vec{x}_T)$ and outputs a sequence of high-level features \vec{h}.
The sequence of high-level features $\vec{h}$ could either be the same length as the acoustic feature sequence or be downsampled if a pyramidal structure is used as in \cite{Chan16}.

The decoder is a stacked unidirectional RNN that computes the probability of a sequence of output units $\vec{y}$ as follows:
\begin{equation}
P(\vec{y}|\vec{x}) = P(\vec{y}|\vec{h}) = \prod_{t=1}^{T}P(y_{t}|\vec{h}, \vec{y_{< t}}) \label{eq:full_decoding}
\end{equation}

At every time step $t$, the conditional dependence of the output on the encoder features $\vec{h}$ is calculated via the attention mechanism. 
The attention mechanism, which is a function of the current decoder
hidden state and the encoder features, condenses the encoder features into a context vector $\vec{c}_{t}$ via the following mechanism:
\begin{align*}
    u_{it} &= \vec{v}^\top \tanh(\vec{W\!_h}\,\vec{h}_i + \vec{W\!_d}\,\vec{d}_t + \vec{b_a}) \\
\vec{\alpha}_{t} & = \text{softmax}(\vec{u}_t) \qquad \vec{c}_t = \sum_{i=1}^{K} \alpha_{it}\vec{h}_{i}
\end{align*}
where the vectors $\vec{v}, \vec{b_a}$ and the matrices $\vec{W_h}, \vec{W_d}$
are learnable parameters; $\vec{d}_t$ is the hidden state of the decoder at time
 step $t$.

The hidden state of the decoder, $\vec{d}_{t}$, which captures the previous output context $\vec{y_{< t}}$, is given by:
\begin{equation*}
\vec{d}_{t} = \text{RNN}(\tilde{\vec{y}}_{t-1}, \vec{d}_{t-1}, \vec{c}_{t-1}) \label{eq:decoder}
\end{equation*}
where $\vec{d}_{t-1}$ is the previous hidden state of the decoder, and $\vec{\tilde{y}}_{t-1}$ is a learned embedding vector for $y_{t-1}$, as is typical practice in RNN-based language models. 
The posterior distribution of the output at time step $t$ is given by:
\begin{equation}
P(y_{t}|\vec{h}, \vec{y_{<\,t}}) = \text{softmax}(\vec{W_\text{s}}[\,\vec{c}_{t}; \vec{d}_{t}] + \vec{b_\text{s}}) \label{eq:posterior}
\end{equation}
where $\vec{W_\text{s}}$ and $\vec{b_\text{s}}$ are again learnable parameters.
The model is trained to minimize the discriminative loss:
$$L_\text{LAS} = -\log(P(\vec{y}|\vec{x}))$$

\subsection{LM Integration Approaches}
Below we discuss the various LM integration approaches for encoder-decoder models that we study.
\subsubsection{Shallow Fusion}
In shallow fusion~\cite{Gulcehre15}, the external LM is incorporated via log-linear interpolation at inference time only.
So while for the baseline model, beam search is used to approximate the solution for:
    $$\vec{y^*} = \argmax_{\vec{y}}\; \text{log}\;p(\vec{y}|\vec{x})$$
    in the most basic version of shallow fusion~\cite{Gulcehre15}, we instead use the following criterion:
    $$\vec{y^*} = \argmax_{\vec{y}}\; \text{log}\;p(\vec{y}|\vec{x})  + \lambda\,\text{log}\;p_{LM}(\vec{y})$$
Recently some additional penalty terms have been introduced in the criterion~\cite{Wu16, Chorowski16, Battenberg17, Kannan18}.  For example, Chorowski and Jaitly~\cite{Chorowski16} use a coverage penalty term $c(\vec{x}, \vec{y})$ to ensure all of the input frames have been ``well attended" during decoding. 

\subsubsection{Deep Fusion}
\begin{figure}[ht]
    \centering
    \includegraphics[height=9cm,keepaspectratio]{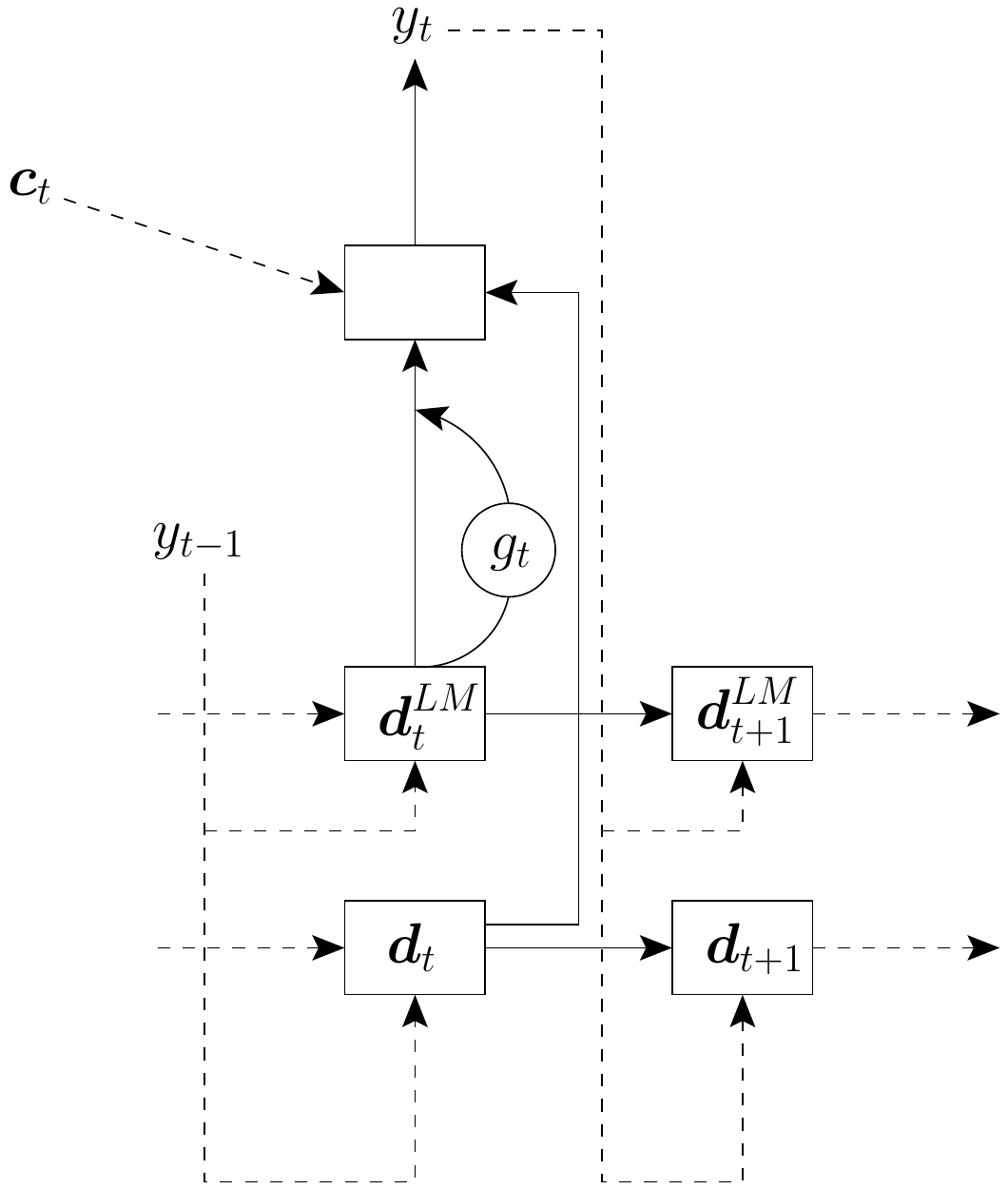}
    \caption{Illustration of a single decoding step of {\em deep fusion}.}
    \label{deep_fusion}
     \vspace{-0.1in}
\end{figure}
Like shallow fusion, deep fusion~\cite{Gulcehre15} 
is a late training integration procedure, i.e.~it assumes the encoder-decoder and language models to be pretrained. 
The key difference is that it integrates the external LM into the encoder-decoder model by fusing together the hidden states of the external LM (assuming a neural LM) and the decoder in the following way:
    \begin{align}
        g_t &= \sigma (\vec{v}_g^T \vec{d}^{LM}_t + b_g)\\
        \vec{d}^{DF}_t &= [\,\vec{c}_{t};\, \vec{d}_t; \,g_t\, \vec{d}^{LM}_t] \label{eq:df_comb}\\
        P(y_t | \vec{h}, \vec{y}_{<\,t}) &= \text{softmax}(\mat{W\!}_{DF\:}\vec{d}^{DF}_t + \vec{b}_{DF})
    \end{align}
where the scalar $b_g$, vectors $\vec{v}_g$ and $\vec{b}_{DF}$, and matrix $\mat{W\!}_{DF}$ are all learned while 
keeping all other model parameters fixed. Fixing most of the model parameters reduces the backpropagation computation cost, and the fine-tuning procedure converges quickly in comparison to the cost of training the baseline model.

\subsubsection{Cold Fusion}
\begin{figure}[ht]
    \centering
    \includegraphics[height=10.5cm,keepaspectratio]{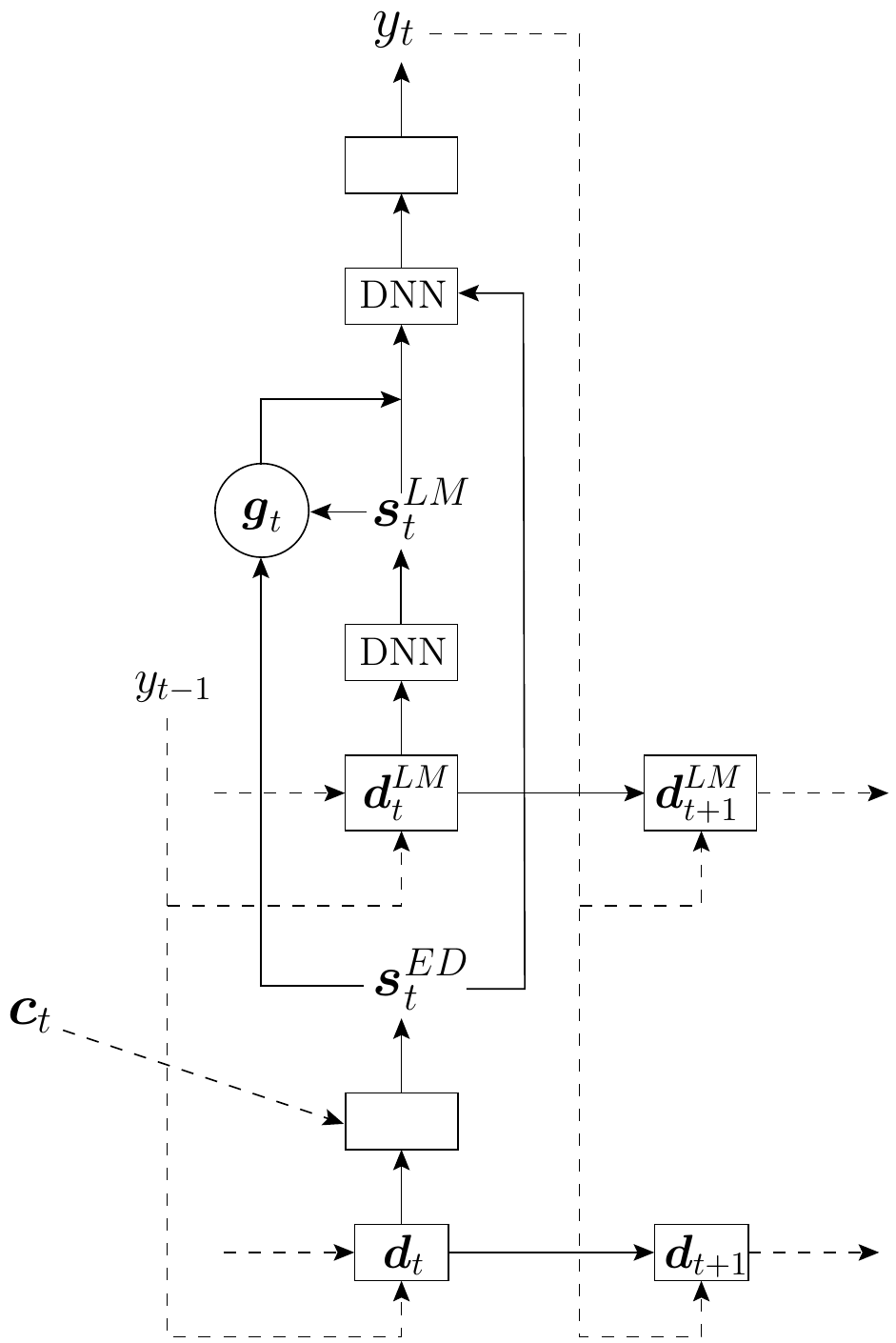}
    \caption{Illustration of a single decoding step of {\em cold fusion}. Note that unlike deep fusion, there is fine-grained gating in cold fusion and hence, gate $\vec{g}_t$ is a vector.}
    \label{cold_fusion}
    \vspace{-0.1in}
\end{figure}
Cold fusion~\cite{Sriram17} 
builds on the idea of deep fusion and proposes a modified LM integration procedure shown below:
    \begin{align}
        \vec{s}^{LM}_t &= \text{DNN}(\vec{d}^{LM}_t) \label{eq:lm_transform}\\
        \vec{s}^{ED}_t &= \mat{W\!}_{ED}\,[\vec{d}_t; \vec{c}_t] + \vec{b}_{ED}\\ 
        \vec{g}_t &= \sigma (\mat{W}_g[\vec{s}^{ED}_t; \vec{s}^{LM}_t] + \vec{b}_g) \label{eq:cf_gate}\\
        \vec{s}^{CF}_t &= [\,\vec{s}^{ED}_t; \;\vec{g}_t \circ \vec{s}^{LM}_t] \label{eq:cf_comb}\\
        \vec{r}^{CF}_t &= \text{DNN}(\vec{s}^{CF}_t) \label{eq:cf_dnn}\\
        P(y_t | \vec{h}, \vec{y}_{<\,t}) &= \text{softmax}(\mat{W\!}_{CF\:}\vec{r}^{CF}_t + \vec{b}_{CF})
    \end{align}
where all of the parameters introduced in the above equations are learned. Some of the key differences between cold fusion and deep fusion are:
\begin{enumerate}[label=(\alph*)]
    \item Cold fusion is an {\em early training integration} approach: The encoder-decoder model is trained from scratch with a pretrained external LM\footnote{LM parameters are kept fixed.}. 
    \item Both the LM state $\vec{s}^{LM}$ and 
    encoder-decoder model's state $\vec{s}^{ED}_t$ are used in gate computation as shown in equation~\ref{eq:cf_gate}.
    \item Cold fusion uses a fine gating mechanism, equation~\ref{eq:cf_comb}, in comparison to a coarse gating mechanism used by deep fusion, equation~\ref{eq:df_comb}.
    \item As originally proposed, cold fusion uses the LM logits rather than the LM hidden state, in order to allow for flexible LM swapping. 
    That is, $\vec{d}^{LM}_t$ used in equation~\ref{eq:lm_transform} refers to the logit scores of the LM rather than the hidden state of LM in the proposed version of cold fusion. 
    However, in practice, with wordpieces used as output units, the relatively large vocabulary results in a long vector of logits $\vec{d}^{LM}_t$  which causes an unnecessary increase in the number of parameters\footnote{The cold fusion paper~\cite{Sriram17} experiments with character level models.}. 
    In our experiments we are not concerned with the flexibility of swapping LMs.
    Hence, in our experiments we still set $\vec{d}^{LM}_t$ to the LM hidden state
    \footnote{Our preliminary experiments with Switchboard suggest a performance gain with this proposed modification.}
\end{enumerate}
Note that, since cold fusion is an {\em early training integration} approach, in a dynamic setting with frequent changes of LM and ASR models the approach would be computationally costlier than the previous two fusion approaches, especially shallow fusion. 

\subsubsection{LM as lower decoder layer}
\label{sec:pretrain}
Previous work in machine translation has suggested the utility of using a pretrained LM as a lower layer of the  decoder~\cite{ramachandran2017unsup}. Similarly, \cite{Rao17} used a pretrained LM to initialize the decoder in an RNN transducer model for speech recognition.
The motivation for this approach 
is that it can provide better contextualized word embeddings, as is the case with the recently proposed Embeddings from Language Models (ELMo)~\cite{peters2018elmo}.
We propose introducing the external LM as a lower layer in the decoder of a pretrained LAS model. All of the model parameters, including the LM parameters,  are fine-tuned for a few epochs with the LAS objective. 

\subsubsection{LM integration via multitask learning}
\label{sec:multitask}
Going back to equation~\ref{eq:full_decoding} of the decoder:
\begin{equation*}
P(\vec{y}|\vec{x}) = P(\vec{y}|\vec{h}) = \prod_{t=1}^{T}P(y_{t}|\vec{h}, \vec{y_{< t}})
\end{equation*}
the decoder can be seen as a conditional LM, conditioned on the encoder features that represent the speech input.
The exact dependence of the decoder on the speech features is captured by the context vector $\vec{c}_t$ which, from equation~\ref{eq:posterior}, affects the output posterior distribution as follows:
\begin{equation*}
P(y_{t}|\vec{h}, \vec{y_{< t}}) = \text{softmax}(\vec{W_\text{s}}[\vec{c}_{t}; \vec{d}_{t}] + \vec{b_\text{s}})
\end{equation*}
Now, unpaired text has no corresponding speech signal.  In the LAS model, this can be represented by a zero context vector.
A zero context vector reduces the decoder from a conditional LM to a plain LM as shown below
\footnote{Note that an ``equivocal" $\vec{c}_t$ that equally affects all of the logit scores 
could also work.  However, such a vector would depend on $\vec{W_\text{s}}$, whereas $\vec{c}_t = \vec{0}$ 
is independent of $\vec{W_\text{s}}$.}:
\begin{equation*}
    P(y_{t}|\xcancel{\vec{h}}, \vec{y_{<\,t}}) = \text{softmax}(\vec{W_\text{s}}[\cancelto{\vec{0}}{\vec{c}_{t}}; \vec{d}_{t}] + \vec{b_\text{s}})
\end{equation*}
In this way, the decoder can also be used for the task of language modeling. Based on this observation, we propose a multitask learning approach for using the unpaired text, where the decoder of the LAS model is shared for the primary ASR task and the auxiliary LM task.
In each iteration of multitask learning, we sample one of the tasks among the ASR and LM task based on the prior probability for picking each task.
Note that when the decoder is trained for LM, the encoder and attention components of LAS model are unaffected.
One important aspect to note is that, unlike all of the previous approaches discussed, this approach has {\em no external LM}; rather, the decoder itself is trained for both tasks.

\section{Experimental Setup}
\subsection{Switchboard}
\subsubsection{Data}
We use the Switchboard corpus (LDC97S62)~\cite{Godfrey92}, which contains roughly 300 hours of conversational telephone speech as our choice of medium-scale training set. 
The first 4K utterances from the training set are reserved as validation set for hyperparameter tuning and early stopping. 
Since the training set has a large number of repetitions of short utterances (“yeah”, “uh-huh”, etc.), we remove duplicates beyond a count threshold of 300. 
After these preprocessing steps, the final training set has about 192K utterances. 
For evaluation, we use the HUB5 Eval2000 data set (LDC2002S09), that consists of two subsets: Switchboard (SWB), which is similar in style to the training set, and CallHome (CH), which contains unscripted conversations between close friends and family.
For acoustic features, we use 40-dimensional log-mel filterbank features along with their deltas, with per-speaker mean and variance normalization. 
For all of the above data processing, we use the EESEN toolkit's recipe~\cite{eesen} which is based on the Kaldi toolkit's recipe~\cite{kaldi}.

For external LM training, we combine the Switchboard training set with the Fisher corpus (LDC200\{4,5\}T19)~\cite{fisher}.
To avoid domain mismatch, we process Fisher utterances to (a) remove noise/hesitation markers not used in Switchboard, and (b) filter out utterances not covered by the wordpiece model trained on Switchboard \footnote{Some utterances in Fisher have symbols such as period sign which are not present in Switchboard and hence are not covered by the wordpiece model trained on Switchboard transcripts.}.
The filtering process removes $\sim$400K utterances out of 2.2 million Fisher utterances.
Thus, combined with the Switchboard training utterances, the LM is trained on $\sim$2 million utterances.

\subsubsection{Model Details}
The encoder is a 4-layer pyramidal bidirectional long short-term memory (LSTM) network~\cite{hochreiter1997lstm}, resulting in an 8-fold reduction in time resolution. 
For the 2-fold reduction done at each layer below the topmost, we max-pool over 2 consecutive hidden states and feed the result into the layer above.
We use 256 hidden units in each direction of each layer.

The decoder in the baseline LAS model is a single-layer unidirectional LSTM network with 256 hidden units. 
We use a 1K wordpiece output vocabulary, which includes all the characters to ensure open-vocabulary coverage. The vocabulary is generated using a variant of the byte pair encoding (BPE) algorithm~\cite{Sennrich16} implemented in the SentencePiece library by Google\footnote{\url{https://github.com/google/sentencepiece}}.
We represent the wordpieces with 256-dimensional embeddings learned jointly with the rest of the model.
For regularization we use: (a) {\em label smoothing}~\cite{Szegedy15},  
where we uniformly distribute 0.1 probability mass among labels other than the ground-truth label, 
and (b) {\em dropout}~\cite{rnn_dropout} with probability 0.1 applied on outputs all of the RNN layers. 
We also use scheduled sampling~\cite{sampling} with a fixed schedule, where each timestep's decoder input is either the ground-truth previous label with probability 0.9 or sampled from the model's posterior distribution for the previous label with probability 0.1. 

For inference, we use beam search with beam size of 10.
We observed that for some of the models increasing the beam size to 10 resulted in escalation of insertion errors compared to a lower beam size. 
To counter this, we add a wordpiece insertion reward $\in \{0, 0.1, 0.2, 0.3, 0.4, 0.5, 0.6, 0.7, 0.8, 0.9\}$, tuned on the development set.
With the addition of the wordpiece insertion reward, larger beam sizes outperform smaller beam sizes for all models, with insignificant gains beyond a sufficiently large beam size of 10.
For shallow fusion, we pick the LM weight $\lambda$ from $\{0, 0.05, 0.1, 0.15, $\\$0.2, 0.25\}$ by tuning on the development set, resulting in a final tuned value of $\lambda=0.2$.

The external LM is a single-layer 512 hidden unit RNN with LSTM cells.
The RNN hidden state is first passed through a projection layer with 256 hidden units and finally fed into the softmax layer.
The LM is trained with the same output vocabulary as the LAS model. 
The LM is trained for 20 epochs with early stopping based on development set perplexity and attains a perplexity of $\sim$15 on the Switchboard development set.

\subsubsection{Training Details}
We bucket our training data by utterance length into 5 buckets, restricting utterances within a minibatch to come from a single bucket for training efficiency.
Different minibatch sizes are used for different buckets, with a batch size of 128 used for the shortest utterances and a 32 batch size used for the bucket with longest ones. 
Preliminary experiments suggested a performance benefit by proceeding through the training set from the bucket with smallest utterances to the one with longest utterances in each epoch.
We use this order for training all of our models.  (A similar training order scheme was also used in~\cite{Audhkhasi18}.)

All models are trained using the Adam optimizer~\cite{adam} with an initial learning rate of 0.001. 
For the baseline LAS model and models with early LM training integration, we train for 12 epochs and start halving the learning rate every epoch after 7 epochs.  
For the models with late LM training integration,  we start halving the learning rate every epoch after 4 epochs and train for a total of 8 epochs.  
For all models, we use early stopping based on development set WER when using greedy decoding.

To speed up training, the encoder of all models is initialized with the encoder of a LAS model trained for predicting phone sequences, similarly to~\cite{Audhkhasi18}.
All models are trained on a single NVIDIA TitanX GPU and finish training within 2 days, with each epoch taking 3-4 hours.  
Finally, all of our models are implemented in TensorFlow~\cite{tensorflow}.

\subsection{Google Voice Search and Dictation}

\subsubsection{Data}

The training data consists of approximately 22 million anonymized, human-transcribed utterances representative of live Google traffic, both Voice Search and dictation.  Clean utterances are artificially corrupted using a room simulator, adding varying degrees of noise and reverberation such that the overall SNR is between 0dB and 30dB, with an average SNR of 12dB. The noise sources are from YouTube and daily life noisy environmental recordings.  The models are evaluated on two data sets: \emph{VS14K}, which consists of about 14K Voice Search utterances, and \emph{D15K}, which contains about 15K dictation utterances.

The external LM is trained on a variety of text data sources, including untranscribed anonymized voice queries (both search and dictation), anonymized typed queries from Google Search, as well as the transcribed training utterances mentioned above.  
Since these component data sources have varying sizes, we up- and down-sample to mix them at a 1:1:1 ratio.

\subsubsection{Model Details}

Our LAS model is consistent with \cite{Chiu18}:  The encoder is composed of 5 unidirectional LSTM layers of 1400 hidden units each, the attention mechanism is a multi-headed additive attention with four heads, the decoder consists of 2 unidirectional LSTM layers of 1024 hidden units each, and the output vocabulary is 16384 wordpieces.  We use 80-dimensional log-mel filterbank features, computed with a 25ms window and shifted every 10ms. Similarly to \cite{sak2015, pundak2016}, at each frame \emph{t}, these features are stacked with 3 frames to the left and downsampled to a 30ms frame rate.

As in \cite{Chiu18}, inference is done via beam search with a beam size of 8.  Shallow fusion numbers are reported after tuning the LM weight $\lambda$ over the values $\{0, 0.05, 0.1, 0.15, 0.2, 0.25, 0.3, 0.35\}$ and a coverage penalty over the values $\{0, 0.01, 0.02, 0.03, 0.04, 0.05\}$, following \cite{Kannan18}. 
These parameters are tuned on a development set consisting of about 10K Voice Search utterances.

The external recurrent LM is composed of 2 LSTM layers of 2048 hidden units each.  It has the same wordpiece output vocabulary as the LAS model. 

\subsubsection{Training Details}

LAS models are trained in two stages.  First, they are trained to convergence with a cross-entropy criterion using synchronous replica training \cite{Chiu18}.  We use tensor processing units (TPUs)~\cite{Jouppi17}
with a topology of 8x8, for a total of 128 synchronous replicas and an effective batch size of 4096.  We found that having a very large batch size was critical to seeing any improvement from cold fusion. Our learning rate schedule includes an initial warm-up phase, a constant phase, and a decay, consistent with \cite{Chiu18}.

Next, we conduct a second training phase with a minimum word error rate (MWER) criterion \cite{Rohit18}.  This phase is performed on 16 synchronous GPU replicas to convergence, which is typically about one epoch.  Note that for deep fusion we effectively have four training phases: cross-entropy training of LAS, MWER training of LAS, cross-entropy training of deep fusion, MWER training of deep fusion.

The external LM is also trained on TPUs with a topology of 4x4.  All models are trained using the Adam optimizer~\cite{adam} and are implemented in TensorFlow~\cite{tensorflow}. 
\section{Results}
\subsection{Fusion Approaches}
\begin{table}[h]
\centering
\caption{Word error rates (\%) on Eval2000 for the baseline model and fusion approaches. SWB=Switchboard, CH=CallHome, Full=Eval2000.}
\begin{tabular}{lccc}
\toprule
Model           &  SWB  &   CH & Full \\
\toprule
LAS        & 17.1   & 27.9    & 22.6    \\
Shallow Fusion  & \textbf{15.6}   & \textbf{26.6}    & \textbf{21.1}    \\
Deep Fusion     & 16.3    & 27.2    & 21.7    \\
Cold Fusion     & 16.3    & 27.3    & 21.8    \\
\bottomrule
\end{tabular}
\label{fusion:swbd}
\end{table}
Table~\ref{fusion:swbd} shows the results of a baseline LAS model and the three fusion approaches on Switchboard and CallHome. 
All of the fusion approaches improve over the baseline model with a relative WER reduction of 3-7\% on Eval2000. Among the fusion approaches, shallow fusion is a clear winner with almost double the gains over baseline compared to deep and cold fusion. Finally, deep and cold fusion have comparable performance on Switchboard. 

\begin{table}[h]
\centering
\caption{Word error rates (\%) on Google voice search (VS14K) and dictation data sets (D15K) for the baseline model and fusion approaches.}
\begin{tabular}{lccc}
\toprule
Model           &  VS14K & D15K    \\
\toprule
LAS        & 5.6   & 4.0
  \\
Shallow Fusion  & \textbf{5.3}	   & \textbf{3.7}
  \\
Deep Fusion     & 5.5    & 4.1 \\
Cold Fusion     & \textbf{5.3}    & 3.9 \\
\bottomrule
\end{tabular}
\label{fusion:google}
\end{table}
Table~\ref{fusion:google} shows the corresponding results for VS14K and D15K. 
All of the fusion approaches improve performance over the baseline model for VS14K, but for D15K deep fusion suffers a minor degradation compared to baseline. 
As with Switchboard, on both of these data sets shallow fusion is again the best performer, although it is tied with cold fusion on VS14K. Finally, deep fusion has no or negligible gain over baseline, suggesting that deep fusion does not scale well with data.

\subsection{Proposed Approaches}
\begin{table}[h]
\centering
\caption{Word error rates (\%) on Eval2000 for the proposed approaches.}
\begin{tabular}{lccc}
\toprule
Model           &  SWB  &   CH & Full \\
\toprule
LM multitask                 & 17.0   & 27.5    & 22.3    \\
Additional layer decoder &     &     &                          \\
\hspace{0.15in} Random initialization   & 16.7    & 28.0    & 22.4           \\
\hspace {0.15in} LM pretrained          & 16.3    & 27.2   & 21.8        \\

\bottomrule
\end{tabular}
\label{prop:swbd}
\end{table}

Next we present results of our proposed approaches (Section~\ref{sec:pretrain} and \ref{sec:multitask}) on Switchboard in Table~\ref{prop:swbd} and compare them against the earlier Switchboard results from Table~\ref{fusion:swbd}. 
The multitask learning approach achieves minor gains over the LAS baseline.  While these gains are more modest than those of the three fusion approaches, it is important to note that unlike the fusion approaches, the LM multitask approach introduces no new parameters. 

Next we evaluate the approach of introducing the LM as a lower decoder layer (making the decoder two layers deep in the case of our Switchboard models). To account for the confounding variable of a deeper decoder, we also compare it to a version where a randomly initialized RNN is introduced instead of the pretrained RNN LM.
As can be seen from the table, the performance of this approach is comparable to that of deep and cold fusion.  In addition, the marginal gains from introducing a randomly initialized RNN instead demonstrate the benefit of LM pretraining.
The promising performance we see here is consistent with the findings of \cite{Rao17} using a pretrained LM as the decoder, and this simple approaches of using a LM to initialize parts of the decoder warrants further investigation in future work.

\subsection{Second Pass Rescoring}
\begin{table}[h]
\centering
\caption{Word error rates (\%) for rescoring on Google data sets.}
\begin{tabular}{lccc}
\toprule
Model           &  VS14K (oracle) & D15K (oracle)   \\
\toprule
LAS        & 5.4 (2.2) & 3.9 (1.5) \\
Shallow Fusion  & 5.3 (2.4) & \textbf{3.7} (1.6) \\
Deep Fusion     & 5.4  (2.0) & 4.0 (1.5)  \\
Cold Fusion     & \textbf{5.0} (1.8) & 3.8 (1.2) \\
\bottomrule
\end{tabular}
\label{second_pass}
\end{table}

While shallow and cold fusion have very similar top-1 WER on the Google data sets, we can investigate the quality of the top-8 to better understand the
strengths of each approach.  For each of the fusion methods, Table \ref{second_pass} shows the WER after second pass rescoring with a large,
production-scale LM (as used in \cite{Chiu18}), as well as the oracle WER in parentheses.

As the table shows, cold fusion has significantly better oracle WER on
\emph{VS14K} than the baseline and other fusion methods and,
as a result, benefits the most from a second pass LM.
While shallow fusion is unaffected by the second pass, the cold fusion WER drops from
5.3 to 5.0.  This suggests that the improvements provided by shallow fusion
are redundant with the benefits of second pass rescoring, whereas cold fusion
does something distinct, improving the overall quality and diversity of the
top 8 decoded transcripts.

Cold fusion also has the lowest oracle WER on \emph{D15K}, but none of the
models benefit much from the second pass on this data set.  The lack of improvement is likely because the second pass
LM is primarily designed to improve performance on Voice Search.  Shallow fusion therefore remains
best on this data set.

Finally, we note that shallow fusion actually has higher oracle WER than the
baseline LAS system on both data sets.  This may be because shallow fusion
can actually pull poor transcripts into the beam if they are heavily
favored by the LM.
\section{Conclusion}
We perform a thorough investigation of the problem of LM integration in encoder-decoder based ASR models.
We compare some of the most prominent past methods and a few of our own proposed methods on the medium-scale and publicly available Switchboard dataset and the large-scale Google voice search and dictation data sets.
Our results show that for first-pass scoring, the simple approach of shallow fusion performs best on all of our data sets. 
However, cold fusion produces lower oracle error rates among the top-8 decoded transcripts, and outperforms shallow fusion after second pass rescoring on Google voice search.
Deep fusion is comparable to cold fusion on Switchboard but gets no or negligible gains over the baseline on Google data sets, suggesting that it does not scale well with data.
Among our proposed methods, the simple approach of using a pretrained language model as a lower layer of the decoder performs comparably to cold and deep fusion on Switchboard, suggesting that further investigation of the approach may be fruitful.

\bibliographystyle{IEEEbib}
\bibliography{references}

\end{document}